\documentclass[12pt]{article}
\usepackage{amsmath}
\usepackage{graphicx,psfrag,epsf}
\usepackage{enumerate}
\usepackage{natbib}
\usepackage{hyperref}
\newcommand{\blind}{0}

\usepackage{graphics}

\usepackage{adjustbox}
\usepackage{amssymb}
\usepackage{bm}
\usepackage{graphicx}
\usepackage{natbib}
\usepackage{amsmath}
\makeatletter
\newcommand{\distas}[1]{\mathbin{\overset{#1}{\kern\z@\sim}}}%
\newsavebox{\mybox}\newsavebox{\mysim}
\newcommand{\distras}[1]{%
  \savebox{\mybox}{\hbox{\kern3pt$\scriptstyle#1$\kern3pt}}%
  \savebox{\mysim}{\hbox{$\sim$}}%
  \mathbin{\overset{#1}{\kern\z@\resizebox{\wd\mybox}{\ht\mysim}{$\sim$}}}%
}
\usepackage{subcaption}
\usepackage{breqn}
\usepackage{multirow}
\usepackage{bm}
\usepackage{multirow}
\usepackage[normalem]{ulem}
\useunder{\uline}{\ul}{}
\usepackage{graphicx}
\usepackage{wrapfig}
\usepackage{lscape}
\usepackage{rotating}
\usepackage{epstopdf}
\usepackage{comment}
\usepackage{xcolor}

\usepackage[utf8]{inputenc}
\usepackage[english]{babel}
 \usepackage{amsthm}

\usepackage{graphicx}

\usepackage{booktabs,amsfonts,dcolumn}

\usepackage{comment}

\begin{document}

\def\spacingset#1{\renewcommand{\baselinestretch}%
{#1}\small\normalsize} \spacingset{1}


\if0\blind
{
  \title{\bf A Space-time Model for Inferring A Susceptibility Map for An Infectious Disease} 

\author{Xiaoxiao Li$^{1}$, Matthew Ferrari$^{2}$, Michael J. Tildesley$^{3}$, Murali Haran$^{1}$\footnote{Email: mharan@stat.psu.edu}}

    \date{
    \small{$^{1}$Department of Statistics, Pennsylvania State University, University Park, PA, USA\\ $^{2}$Department of Biology and the Center for Infectious Disease Dynamics, Penn State University, University Park, PA, USA\\ $^{3}$The Zeeman Institute for Systems Biology and Infectious Disease Epidemiology Research,
School of Life Sciences and Mathematics Institute, University of Warwick, Coventry, UK}}
  \maketitle
} \fi

\if1\blind
{
  \bigskip
  \bigskip
  \bigskip
  \begin{center}
    {\LARGE\bf Title}
\end{center}
  \medskip
} \fi

\bigskip
\begin{abstract}
Motivated by foot-and-mouth disease (FMD) outbreak data from Turkey, we develop a model to estimate disease risk based on a space-time record of outbreaks. The spread of infectious disease in geographical units depends on both transmission between neighbouring units and the intrinsic susceptibility of each unit to an outbreak. Spatially correlated susceptibility may arise from known factors, such as population density, or unknown (or unmeasured) factors such as commuter flows, environmental conditions, or health disparities. Our framework accounts for both space-time transmission and susceptibility. We model the unknown  spatially correlated susceptibility as a Gaussian process. We show that the susceptibility surface can be estimated from observed, geo-located time series of infection events and use a projection-based dimension reduction approach which improves computational  efficiency. In addition to identifying high risk regions from the Turkey FMD data, we also study how our approach works on the well known England-Wales measles outbreaks data; our latter study results in an estimated susceptibility surface that is strongly correlated with population size, consistent with prior analyses. 

\end{abstract}

\noindent%
\begin{center}
    Foot and Mouth Disease, Gaussian process, dimension reduction
\end{center}

\spacingset{1.5}
\newpage
\section{Introduction}
\label{sec:intro}

With the availability of 
geo-locational infectious disease surveillance data, epidemiologists are able to better study  transmission patterns and disease dynamics through spatio-temporal modeling, which not only have public health impacts but also wide-ranging socio-economic influence \citep{lessler2016mechanistic, smith2019infectious}. 
To better provide disease control guidance and achieve fast reaction towards potential outbreaks, it is important to pay attention to the spatially varying underlying risks of susceptibility  that are usually overlooked in favor of the space-time transmission behavior alone. In this paper we propose a unit-level space-time model of infectious disease taking into account both space-time disease transmission dynamics and a spatially varying underlying susceptibility. Here, susceptibility describes how severely a unit is influenced by the transmission force that may have originated from the surrounding infectious units. Susceptibility can hence also be thought to account for factors associated with a unit that are hard to quantify or not always available such as commuter flows, environmental conditions, population or health disparities. 
Susceptibility can be treated as a proxy for how likely a unit is to be infected if exposed to infected neighbors, thus one direct application of our model is to produce maps of susceptibility risk, which can help identify spatial clusters of high risk, reveal public health challenges and provide an essential evidence base to guide policy decisions in a visually condensed form \citep{reich2018precision}.

To characterize the transmission mechanism of infectious disease epidemics in time and space,  numerous space-time models have been proposed and applied to solve ecological and public health questions, for example, transmission networks models \citep{garnett1996sexually,keeling2005networks}, deterministic continuous-space models \citep{noble1974geographic,murray1986spatial,caraco2002stage} and stochastic metapopulation models \citep{hanski1991metapopulation}. 
In this paper, our infectious disease model builds upon the Warwick model transmission framework \citep{keeling2001dynamics, tildesley2006optimal, deardon2010, lawson2013statistical} which has been widely used to provide an intuitive and flexible framework to study foot-and-mouth disease. These individual-based models quantify the probability of a susceptible individual being infected from the surrounding infectious pressure while considering individual-level heterogeneity by taking local demographic factors such as livestock numbers and species into account.  
The quality of surveillance data may be highly variable due to differences in investment, planning, resources, and specific characteristics of the disease and the region of study. Sparse surveillance data and lack of locally specific demographic data present a significant hurdle to the development of control and eradication programs in new areas \citep{thacker1983surveillance}. Our approach provides some potential ways to address these issues. Our model 
only requires geo-located time series records of outbreaks, where outbreaks are binary values that indicate whether there is an outbreak at that location at the time, to infer the susceptibility of a given unit. We will refer to our geographical units where outbreaks are observed as 'epi-units'. An epi-unit can be a farm, village, district or even country depending on the spatial resolution of the data.  Gaussian process structure provides a flexible stochastic model, which can capture spatial correlations in random effects as well as disease-driving factors such as the environment through spatial varying coefficients \citep{gelfand2003spatial}. 
Furthermore, the model allows us to interpolate to units where surveillance data may be missing. For units outside of the surveillance system, our models are able to borrow information from neighboring units by imposing a Gaussian process structure on the susceptibilities we are interested in inferring. While these models are very flexible, the large number of highly correlated spatial random effects result in costly likelihood evaluations and slow mixing in Markov chain Monte Carlo (MCMC) algorithms \citep{christensen2006robust, guan2018computationally}. We use a projection-based intrinsic conditional autoregression
(PICAR) approach, which is a discretized and dimension-reduced representation of
the underlying spatial random field using empirical basis functions on a triangular
mesh \citep{lee2019picar} to provide a fast computational approach for our model. 

Our contribution may be summarized as follows: (1) We propose an infectious disease model that  accounts for both space-time transmission and the underlying susceptibility; the inferred susceptibility map can provide guidance for learning about the mechanisms that give rise to the susceptibilities. (2) Our modeling framework allows us to include spatial dependence among susceptibilities along with a heuristic for determining when spatial dependence should be incorporated. (3) We apply dimension reduction techniques on spatial varying coefficients to achieve computational efficiency.


The rest of the paper is organized as follows. In Section 2 we introduce foot-and-mouth disease (FMD) surveillance data in Turkey which act as motivation for the development of our model. We also briefly describe the pre-vaccination era measles data from England and Wales. In section 3 we present our modeling framework and in Section 4, we describe details on how we carry out inference. In Section 5, we study our approach via extensive simulations and present the results of applying our models to two real data sets: FMD outbreaks in Turkey from 2001 to 2012 and the well-studied measles surveillance data in England and Wales. In Section 6, we discuss the range of applications for our model and some directions for future study and research.

\section{Data}
In this section we introduce two  data sets to which we will apply our methods. The first is a foot-and-mouth disease surveillance. 
The second is measles outbreak data in England and Wales in the pre-vaccine era.

\subsection{Foot-and-mouth Disease in Turkey}
Foot-mouth disease (FMD) is a  severe and highly infectious viral disease that affects cloven-hoofed ruminants, and has become a worldwide concern for livestock production since the beginning of the 20th century \citep{Grubman465}. Although FMD does not cause high mortality,  infected animals suffer from reduced milk yield, weight loss and lameness, resulting in a decrease  in productivity for a considerable amount of time. 
In many lower and middle income countries in Africa, the Middle East and Southern Asia, where a significant proportion of the population rely upon livestock for their livelihoods, regular waves of FMD outbreaks affect the majority of households directly. In terms of visible production losses and vaccination costs, the annual economic impact of FMD  in endemic regions of the world is estimated to be between USD 6.5 and 21 billion; outbreaks in FMD‐free countries and zones cause losses of more than USD 1.5 billion per year \citep{economic}. 
\begin{figure}
    \centering
    \includegraphics[width= 0.8\textwidth, height = 0.4\textwidth]{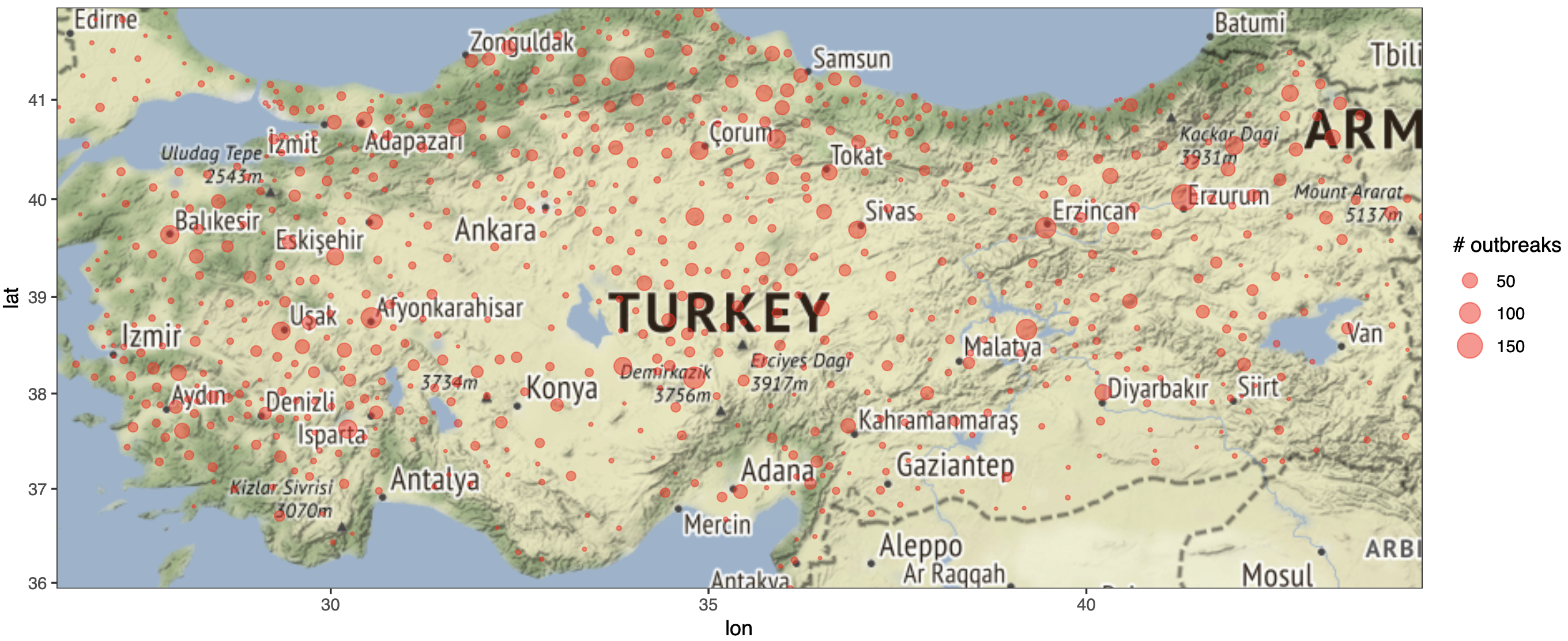}
    \caption{Number of outbreaks per district in Turkey between 2001-2012.}
    \label{fig:count_map}
\end{figure}

A significant decrease in FMD incidence from the 1970s to 2000 in Turkey is coupled with a rise in vaccine production and attention to national campaigns; however the disease continues to persist in the country and the dynamics of each serotype group appear to be distinct \citep{FMDpattern}. Considerable efforts in management and surveillance have been made and local control measures are implemented following any reported outbreaks. These include short-term movement restrictions, active surveillance to improve detection, and vaccination of all susceptible cattle herds within a 10 km radius. Evidence suggests that the efficacy of the vaccines in use is relatively low, resulting in protection of around 50\% of the cattle population \citep{50vaccine}, and limited resources can present a challenge for the successful implementation of surveillance and movement control. In addition, there appears to be little to no protection between FMD serotypes \citep{noCrossImmune} indicating that animals that have recovered from infection with one serotype may remain susceptible to infection from others.

\begin{figure}
    \centering
    \includegraphics[width= 0.8\textwidth, height = 0.4\textwidth]{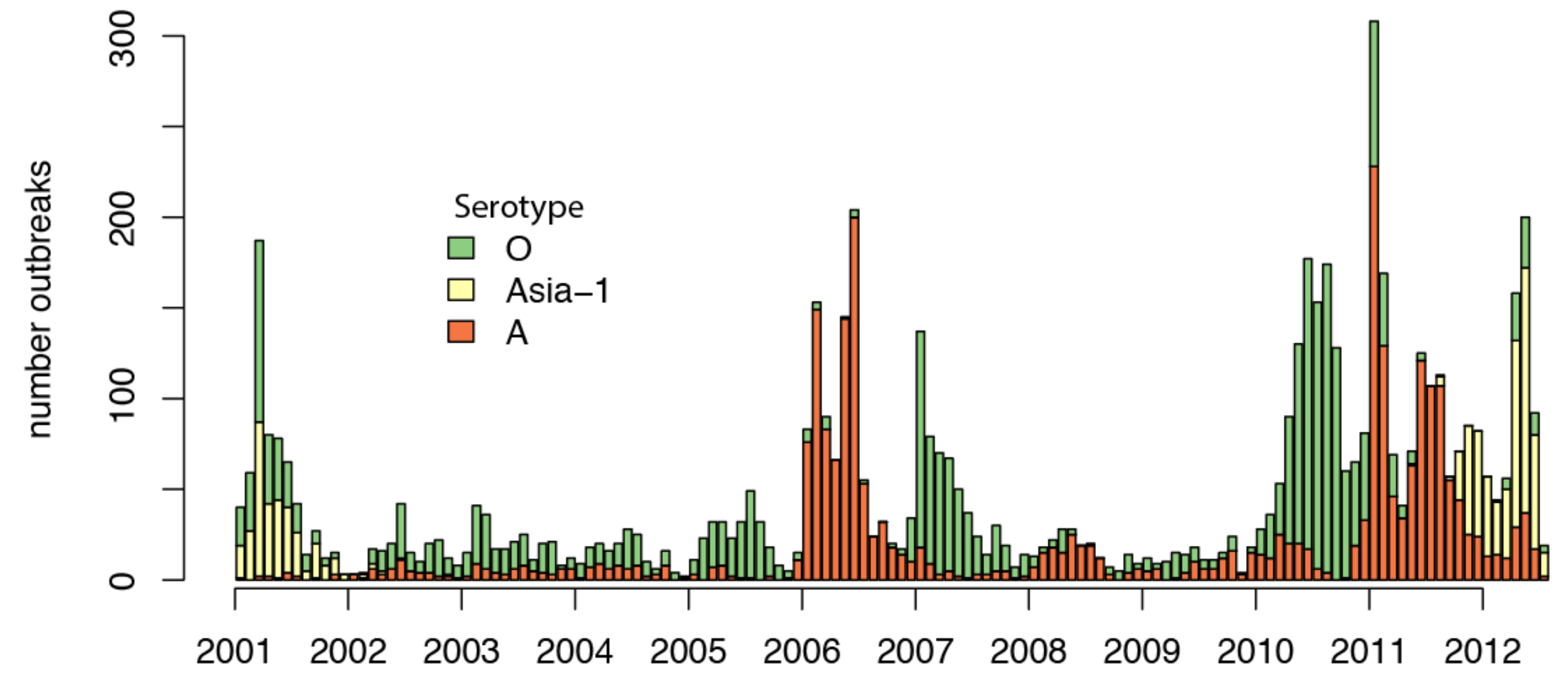}
    \caption{Number of outbreaks per month attributed to each serotype between 2001-2012.} 
    \label{fig:strain_ts}
\end{figure}

Within Turkey there are distinct geographical differences in FMD prevalence. Thrace, in the west of the country, has been FMD free with vaccination since 2010. 
However, in the remainder of the country, there is continual circulation of multiple serotypes. Serotype O is the most dominant strain and persists at a relatively high incidence, followed by serotype A which is observed at a lower incidence. The Asia-1 serotype was largely unobserved in the country during the first decade of the 21st century, before re-emerging in the east of the country in 2011 in some provinces (Figure \ref{fig:strain_ts}). 

Between 2001 and 2012, there were 9282 FMD outbreaks at  epi-unit (approximately village) level. For each outbreak, we have records of the location to the epi-unit level (Figure \ref{fig:count_map}, aggregated to district for visualization), start/end dates, and partial information on total livestock, numbers of animals infected and culled, and date of culling. In addition, the serotype was identified in 70\%  of outbreaks (Figure \ref{fig:strain_ts}). These surveillance information are aggregated to epi-unit level \citep{dawson2016analysis} and enable us to build a mathematical model to evaluate the risk for multiple serotypes at various time interval of interests in Turkey.

\subsection{Measles in England and Wales}

We also study the weekly reports of measles cases for 952 municipalities in England and Wales in the pre-vaccination time from 1944 to 1965. This dataset is well-studied \citep[cf.][]{grenfell2002measlesdynamics, xia2004measles, jandarov2014emulating} and strong correlation has been detected between disease risk and population \citep{keeling1997disease, bjornstad2008hazards}. By applying our proposed model to these data, we are able to see whether our model reveals the susceptibility of different municipalities to an outbreak in a manner that is consistent with previous studies. 

\section{Methods}

Motivated by Turkey's FMD outbreak surveillance data, we present a model that describes space-time transmission mechanism and provides unit-level (district-level) susceptibility risk  estimation, where the units are represented as discrete points in time and space that records the starting time of the outbreaks that were aggregated and reported at district-level. In this section, we first introduce the Warwick model \citep{keeling2001dynamics} used to describe FMD transmission dynamics. We then describe our model, which builds upon the Warwick model and makes it more suitable for the aggregate level data we are confronted with in Turkey. Our model provides a framework for estimating both transmission dynamics as well as potentially spatially dependent susceptibility of each location to an outbreak. 


\subsection{Modeling Framework}

The mathematical model for describing transmission behavior for FMD was first proposed by  \cite{keeling2001dynamics} to predict spread and optimal control strategies for the 2001 FMD outbreak in the UK. The model has been adapted to various scenarios both for the UK \citep{tildesley2006optimal, deardon2010} and elsewhere in the world such as Denmark \citep{tildesley2008modelling} and the USA \citep{tildesley2012modeling}. We assume the surveillance data provides us accurate information about the status of infection for all the 957 districts, i.e., susceptible or infected, for each time-step which may be aggregated to week or month to avoid sparsity. Since numerous consecutive infections in time are observed in one unit, we adopt a Susceptible-Infectious-Susceptible (SIS)  model to characterize the transmission behavior between units and assign compartments for every unit at each time-step accordingly. 

Following the model framework from \cite{keeling2001dynamics} on unit level outbreaks, the rate at which a susceptible unit\footnote{For unit that are already infected at time $t$, there is no need to describe its risk of infection.} $i$ gets infected at time $t$ due to transmission depends on all the units that were in infected status at time $t-1$ which are contained in the infected compartment  $I_{t-1}$:

\begin{equation}\label{e1}
    \lambda_i(t)= \beta_i\times \sum_{j \in I_{t-1}}  k(d_{i,j}),
\end{equation}
where $k(d_{i,j})$ is a transmission kernel which is a decreasing function of distance that  accounts for risk of infection, where $d_{i,j}$ represents the Euclidean distance between unit $i$ and unit $j$ (some other distance measures are also considered in  \citeauthor{savill2006topographic} \citeyear{savill2006topographic}). The probability that a susceptible unit $i$ is infected at time $t$ is represented as 
\begin{equation}\label{e2}
P_i(t) = 1- \exp\left\{- \lambda_i(t) - \gamma\right\},
\end{equation}
where $\gamma$ is the background spontaneous rate of infection that captures any factors other than transmission from the neighboring units. 
Because of data quality limitations, we choose not to include an extra term for number of livestock; instead, by using a random susceptibility term $\beta$, we attempt to capture the impact of the number of livestock, along with other unobserved characteristics of the epi-unit, on susceptibility. In addition, one of our goals is to have an approach that easily generalizes to places with sparse or missing unit level demographical data; such data quality issues are common in developing countries \citep{jamison2006disease}. As a result, we choose to let susceptibility term $\beta$ to be flexible for each unit. It can be interpreted as a scalar of the transmission force coming to susceptible unit $i$, with larger $\beta$ having higher risk and small $\beta$ with lower risk.

\subsection{Independent Susceptibility  Model (ISM)}
Susceptibility term $\beta$ is a scaling term that impacts the probability of an outbreak given infections in other places. It  can provide guidance in disease control and policy implementation. In our first model we allow $\beta$ to vary independently across epidemiological units, that is. $\beta = [\beta_1,...,\beta_n]$ to account for variations in unknown or unmeasured factors such as commuter flows, environmental conditions, or health disparities.  Let $Y_{i,t} \in \{0,1\}$ be the binary outbreak status of unit $i \in \{1,..., N\}$ at time $t \in \{1,...,T\}$ denoting that an outbreak was observed, and 0 indicating there was no outbreak at that time. We develop a hierarchical model using the terminology of Berliner \citep{berliner2000hierarchical}.  At the observation level, our goal is to model the probability of the outbreak occurrence in space and time using a Bernoulli random variable with probability defined as in equation \ref{e2}. At the process level, we quantify the infection rate as the force of infection scaled by the susceptibility using equation \ref{e1}. The prior model for susceptibility follows an 
exponential distribution. The resulting model is:
        \begin{equation} \label{e3}
            \begin{split}
            Y_{i,t}&\sim \text{Bernoulli}(1- \exp\left\{- \lambda_i(t) - \gamma\right\}), \\
            \lambda_i(t)&= \beta_i \times \sum_{j \in I_{t-1}}  k(d_{i,j}),\\
            k(d_{i,j}) &= (1+\frac{d_{i,j}}{\phi})^{-b_0},\\
            \beta_i &\sim \text{exp}(\alpha),
            \end{split}
        \end{equation}
where $\alpha$ is the mean of the exponential distribution; by using the exponential prior on $\beta$s ensures positive scaling on a span of $(0, +\infty)$. 

\subsection{Spatially Dependent Susceptibility Model (SDSM)}

When there are regional similarities in susceptibility, $\beta_i$'s are spatially correlated. In addition, for units with limited surveillance records, a spatial dependence assumption allows us to borrow information from nearby units when appropriate.  We use a  Gaussian process 
to account for dependence \citep{haran2011gaussian} and an exponential transformation to ensure positive $\beta$, that is,   $\log{\bm{\beta}}\sim\mathcal{N}(\omega,\Sigma)$, where $\Sigma$ is the spatial covariance matrix constructed by a covariance function and $\omega$ is the mean of $\log(\beta)$'s. In this model, we employ an exponential covariance function: $\Sigma^{i,j} =\sigma^2\exp(d_{i,j}/\rho)$ and the full spatial hierarchical model can be expressed as:

        \begin{equation}\label{e4}
            \begin{split}
            Y_{i,t}&\sim \text{Bernoulli}(1- \exp\left\{- \lambda_i(t) - \gamma\right\}), \\
            \lambda_i(t)&= \beta_i \times \sum_{j \in I_{t-1}}  k(d_{i,j}),\\
            k(d_{i,j}) &= (1+\frac{d_{i,j}}{\phi})^{-b_0},\\
            \log{\bm{\beta}} &\sim \text{GP}(\omega, \Sigma).
            \end{split}
        \end{equation}
Our process encourages regional similarity and can be used to find  regions of higher (or lower) susceptibility. This can be useful to policy makers since disease control plans are often designed and executed at a regional scale.

\label{sec:meth}
\section{Inference}
Analysis for our models are conducted in the Bayesian paradigm through a Markov chain Monte Carlo (MCMC) algorithm and was implemented using statistical programming language Nimble\citep{nimble2017}. In this section, we first provide a heuristic test that helps in choosing  between the independent susceptibility and spatially-dependent susceptibility models by analyzing the correlogram of the cross-entropy. We then describe our two-step inference procedure and provide details of our MCMC implementation.

\subsection{Deciding between Independence and Dependence Models}\label{dependence_test}
The main difference between ISM and SDSM is whether there is spatial dependence in conditional risk $\beta$'s. We provide an ad hoc procedure to choose between the two models.
\begin{itemize}
    \item[Step 1]: Fit ISM and obtain estimate for $\beta_1,\dots,\beta_N$. 
\item[Step 2]: Obtain the predicted probability of an outbreak at location $i$ at time $t$:
$$P_i(t) = 1- \exp\left\{- \lambda_i(t) - \gamma\right\}$$
where $\lambda_i(t)$ and $\gamma$ are estimates from Step 1.
\item[Step 3]: Calculate cross-entropy loss \citep{cross-entropy} for each  geographical location by $$L_i = -\sum_{t} \{Y_{i,t}\log(P_i(t)) + (1-Y_{i,t}\log(1-P_i(t))\} $$
\item[Step 4]: Draw a correlogram plot of the residuals to determine whether spatial dependence is significant. 
\end{itemize}

\begin{figure}[!ht]
    \centering
        \includegraphics[height = 0.3\textwidth]{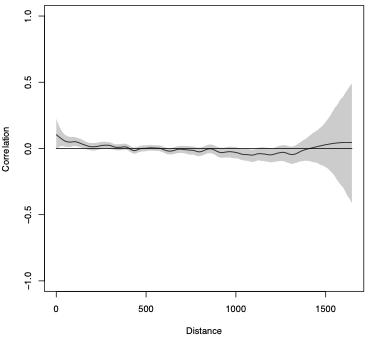}
        \includegraphics[height = 0.3\textwidth]{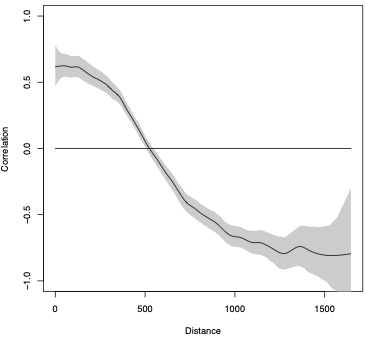}
        \caption{Correlogram of cross-entropy loss of  data simulated with (1) independent susceptibilty (2) spatially dependent susceptibility}
        \label{heuristicplot}
\end{figure}

We use the `correlog' function in `ncf' package \citep{bjornstad2018epidemics} to draw a correlogram with confidence intervals, which  can be used to make a judgement about whether it is important to include spatially dependent susceptibility in the model. We provide an example of this in Figure \ref{heuristicplot}. The correlogram plot shows autocorrelation as a function of distance. In Step 4, if no significant spatial correlation is detected, that is, the estimated correlation function, with error bars, includes 0, ISM with independent relationship in $\beta$'s is recommended; on the other hand, if strong spatial correlation is detected in the cross-entropy loss, it indicates that some other spatial dependence besides transmission is left over and Model II will be a good option.  

\subsection{Two-step Procedure}
To simplify computational challenges that partially arise from the additive relationship between background rate ($\gamma$) and transmission force ($\lambda_i(t)$), we  propose a two-step  estimation approach. We first estimate background rate $\gamma$ and transmission kernel parameter $\phi$ using a combination of method of moments and least squares estimation. In the second step, we fix the estimated $\gamma, \phi$ parameters from the first step and then use a Bayesian approach to infer the  susceptibilities ($\beta$s). 

First, to estimate the background rate $\gamma$, we choose a time window from $t_1$ to $t_2$ that only contains mild outbreak occurrences. Letting $\Delta t = t_2 - t_1$, we obtain a method of moments estimate of background rate, $$\hat{\gamma} = \frac{\#\text{outbreaks} \in (t_1,t_2)}{N\Delta t}.$$ To estimate the parameter $\phi$ of the transmission kernel, we discretize the range of $\phi$  into a sequence as $(\phi_1,...,\phi_m)$. For a given $\phi_k$, we generate the total force of infection at time $t$ as follows: $M_t^k= \sum_{i=1}^N \sum_{j \in I_{t-1}}  k(d_{i,j}) $ for $t=2\dots,T$. For each $\phi_k$, we compute the squared Euclidean distance between this force of infection and the observed epidemic time series. We find our  estimate of  $\phi$ by setting it to the $\phi_k$ that minimizes this Euclidean distance.

After  obtaining the  estimates for $\gamma$ and $\phi$, we proceed with Bayesian estimate of susceptibility $\beta$'s. For models with independent priors on $\beta$ (ISM), MCMC provides a fast estimate with uncertainty represented in the posterior sampling distribution. The SDSM has a large number of highly correlated $\beta$s resulting in costly likelihood evaluation (of order $N^3$) and slow mixing in MCMC algorithms. PICAR \citep{lee2019picar} provides a basis representation of spatial correlation matrix and significantly reduce computational by dimension reduction on susceptibility $\beta$'s . Following the PICAR approach, instead of inferring the covariance matrix directly from $\log{\bm{\beta}} \sim \text{GP}(c_0, \Sigma)$, we represent spatial dependent log susceptibility as a linear combination of basis functions:

\begin{equation}\label{e5}
\begin{split}
    log(\bm{\beta}) &\approx c_0 + \Phi \delta,\\
    \delta &\sim \textbf{N}(0, \Sigma_{\delta}),
\end{split}
\end{equation}

where $\Phi$ is a basis function matrix and $\delta \in \mathbb{R}^p$ are the associated weights. Dimension reduction is achieved by choosing $p$ to be much smaller than  $N$.  In addition,  de-correlation is achieved through orthogonal basis representations which greatly improves the mixing of the MCMC algorithm.

\subsection{MCMC Algorithm}
In this section, we provide the details of MCMC algorithms for fitting the ISM and SDSM models.  

In ISM, we give exponential priors for $\beta$: $\beta \sim \text{exp}(\alpha)$ with $\alpha \sim \text{Uniform}(0,5)$. The full conditiona distriution for $\beta$ is then $$\pi(\beta|.) \propto \prod_{t=1}^T\prod_{i=1}^N \mathit{f} (Y_{i,t}|\hat{\gamma},\lambda_i(t),\beta)p(\beta),$$ where $\lambda_i(t) =\beta_i \times \sum_{j \in I_{t-1}} (1+\frac{d_{i,j}}{\hat{\phi}})^{-b_0}$ and $\mathit{f} (Y_{i,t}|.) = P(Y_{i,t}|.)^{Y_{i,t}}[1-P(Y_{i,t}|.)]^{1-Y_{i,t}} $. 

In SDSM, we structure $\log(\beta) \sim \text{Normal}(\omega, \Sigma)$ where $\Sigma_{i,j} = \sigma^2 \text{exp}(d_{i,,j}/\rho)$ with  a uniform prior on variance parameter $\sigma \sim \text{Uniform}(0,5)$, smoothness parameter $1/\rho \sim \text{Uniform}(0,1)$ and mean parameter $\omega \sim \text{Uniform}(-10,0)$. The full conditional distibution for $\beta$ is then $$\pi(\beta|.) \propto \prod_{t=1}^T\prod_{i=1}^N \mathit{f} (Y_{i,t}|\hat{\gamma},\lambda_i(t),\beta,\omega)p(\beta|\sigma^2, \rho,\omega)p(\sigma)p(\rho)p(\omega).$$

When PICAR is applied in inferring SDSM, instead of updating every $\beta \in \mathbb{R}^N$, our target switch to the weights ($\gamma \in \mathbb{R}^p$) of the Moran's basis function $\mathbf{M} \in \mathcal{R}^{N\times p}$ \citep{hughes2013dimension}, and rely on projection matrix $\mathcal{A}$ \citep{lee2019picar} to recover $\beta$ estimates from the trained basis $\log(\beta) \approx \mathbf{AM}\delta + \omega$. This approach has lower dimension required compared to learning the full N by N covariance matrix $\Sigma$. Following Lee and Haran (2021), we impose Normal priors on the weight of basis $\delta \sim N(0,\tau^{-1}(M^'QM)^{-1})$ where $\tau$ is the precision parameter that follows a gamma distribution: $\tau \sim \text{gamma}(0.5,2000)$, $Q$ is the prior precision matrix for the mesh vertices which is the independent identity matrix. The resulting conditional for $\delta$ is therefore
$$ \pi(\delta|.) \propto \prod_{t=1}^T\prod_{i=1}^N \mathit{f} (Y_{i,t}|\hat{\gamma},\lambda_i(t),\beta(\delta, \omega)) p(\delta|\tau)p(\tau)p(\omega).$$
After we obtained Bayesian estimates of $\delta$ and $\omega$, $\beta$ is recovered through the projection matrix as well through $\log(\hat{\beta}) = AM\hat{\delta} + \hat{\omega}$.

\section{Results}
In this section, we apply our models and computational appproach to real and simulated data sets. The two real data sets -- Turkey FMD surveillance data and England and Wales measles surveillance data -- provide outbreaks that are well documented but have very distinct features, thereby making them useful examples for our study. They allow us to compare the performance of models with different assumptions on susceptibility  and requirements in computational speed. As before, we denote the model with independent susceptibility as ISM, model with spatially dependent susceptibility as SDSM, and SDSM inferred by dimension reduction algorithm PICAR as SDSM PICAR.

\subsection{Simulation}
In order to make the simulations resemble the real data closely, we simulate outbreak data on the map of Turkey at the district level based on the models described in Section 3.  We use an exponential covariance function to describe spatial dependence\footnote{distances are measured in km}, so that the covariance between  susceptibilities at a distance $d_{i,j}$ is  
$\sigma^2_{\beta}\exp(d_{i,j}/\rho),$ with the parameters $\sigma^2, \rho>0$. Here we set  $\sigma^2_{\beta} = 1$, and obtain three levels of spatial dependence by varying $\rho$: low ($\rho = 200$), medium ($\rho = 400$), and high ($\rho = 600$) spatial dependence. In addition to these three  settings, we also explore the case with spatially independent susceptibility. To each of these four different settings, we fit three different models: ISM which assumes independent susceptibility, SDSM which assumes spatially dependent susceptibility, and SDSM-PICAR which is SDSM with a basis representation approach to make the computing more efficient. 

From the simulations we obtain an $N \times T$ matrix of  binary response of outbreaks, where $N$ is the number of districts in Turkey and $T$ is the simulated time steps. We randomly choose 90\% of the locations as training data and treat them as units covered by surveillance; the remaining 10\% of locations are testing data which are analogous to the units outside of the surveillance system.  To evaluate the  performance of the four simulation settings, we train the three models (ISM, SDSM, SDSM-PICAR) based on the training data and report out-of-sample mean squared prediction error (MSPE) 
on the testing set  between the ground truth and the estimates obtained of susceptibility $\beta$. Our study shows that when $\beta$s are spatially independent, ISM outperforms the other two models which assume spatial dependence. For simulation settings with spatially dependent
$\beta$s, models with spatial dependence outperform  ISM. The performance of SDSM and SDSM-PICAR are comparable, with SDSM-PICAR resulting in greatly reduced computational complexity from $O(n^3)$ to $O(np)$. 
The simulation study in Table 1 was set with epidemic length of 100 time steps; we also conduct experiments on time steps of 150 and 200, and similar patterns are also observed. 

Because ordering of risk is of priority from a policy perspective, we examine whether our approach does well in reproducing the ordering of the risks across locations. We found that by using Spearman's correlation, we get a higher correlation between the estimated rank and the ground truth using proposed model than that of the ranking correlation estimated from incidence by location. In the settings from simulation study in Table 1 of 4  levels of spatial dependence (independent, mild, middle, strong), the correlation for the approached based on incidence were 0.790, 0.705, 0.694, 0.887, and for our approach the correlations are 0.813, 0.811, 0.858, 0.927.

\begin{table}
\centering
\begin{tabular}{|l|l|l|l|l|l|}
\hline
 Level of dependence & ISM &  SDSM & SDSM PICAR\\ \hline
  independent& \textbf{0.0572}&  0.0792 & 0.0837\\ \hline
 low & 0.0385 & \textbf{0.0129} & 0.0158 \\ \hline
  medium &0.0270& 0.0138 & \textbf{0.0085} \\ \hline
 high & 0.0692 &  \textbf{0.0145} & 0.0164 \\ \hline
\end{tabular}
\caption{Comparing the performance of different models for simulated examples}
\end{table}

\subsection{Foot-and-mouth Disease Surveillance in Turkey}

Here we apply our methods to an FMD surveillance data set from Turkey, studying susceptibilities to different serotypes. Using the model choice heuristic introduced in Section \ref{dependence_test} we find that there is  considerable spatial autocorrelations within distances of 250 km, which suggests that we include spatial dependence structure in the susceptibility as in the  SDSM model. 
After fitting the models, we further compare SDSM and ISM by a post-analysis in evaluating the out of sample cross-entropy error. We reuse the same approach as Section 5.1 to divide the data between training and testing data. We ind SDSM outperforms ISM again by having smaller cross-entropy loss ($756.6 < 807.9$) on the test data. 

\begin{figure*}
        \centering
        \begin{subfigure}[b]{0.475\textwidth}
            \centering
            \includegraphics[width=\textwidth]{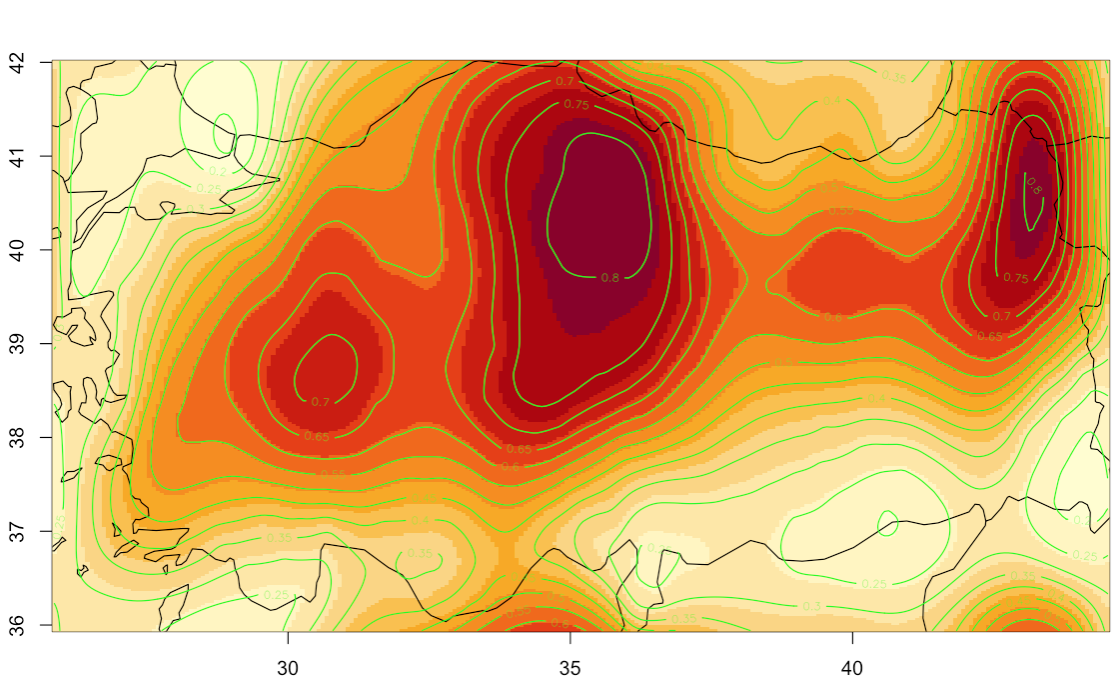}
            \caption[]{}%
        \end{subfigure}
        \hfill
        \begin{subfigure}[b]{0.475\textwidth}  
            \centering 
            \includegraphics[width=\textwidth]{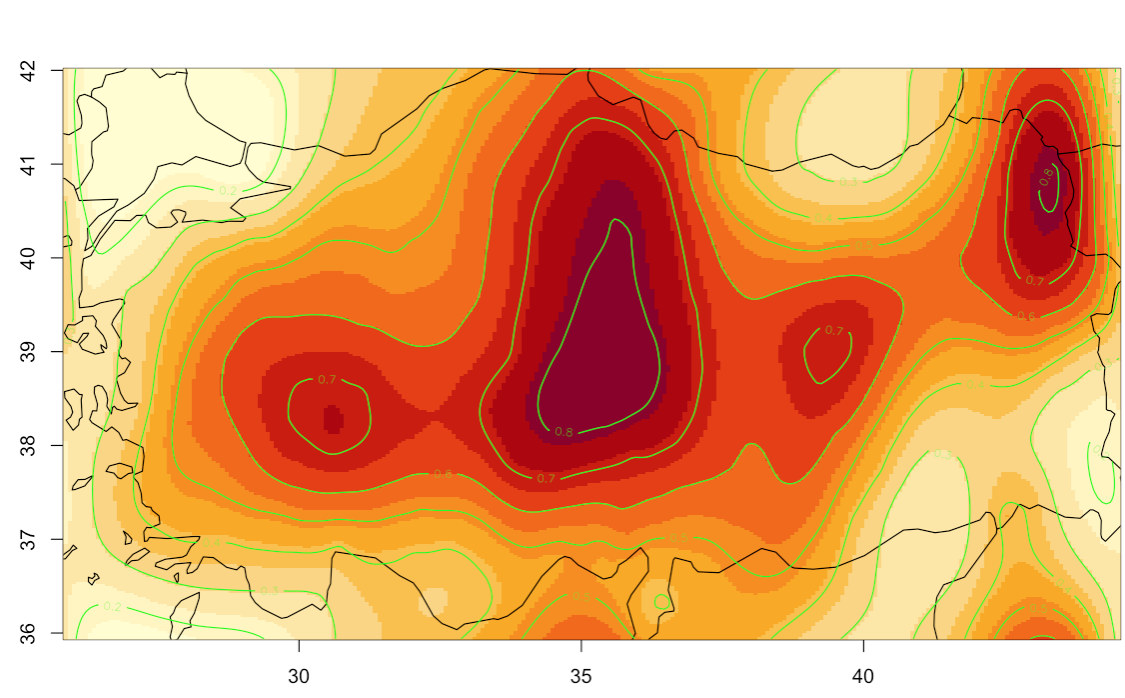}
            \caption[]{}%
        \end{subfigure}
        \vskip\baselineskip
        \begin{subfigure}[b]{0.475\textwidth}   
            \centering 
            \includegraphics[width=\textwidth]{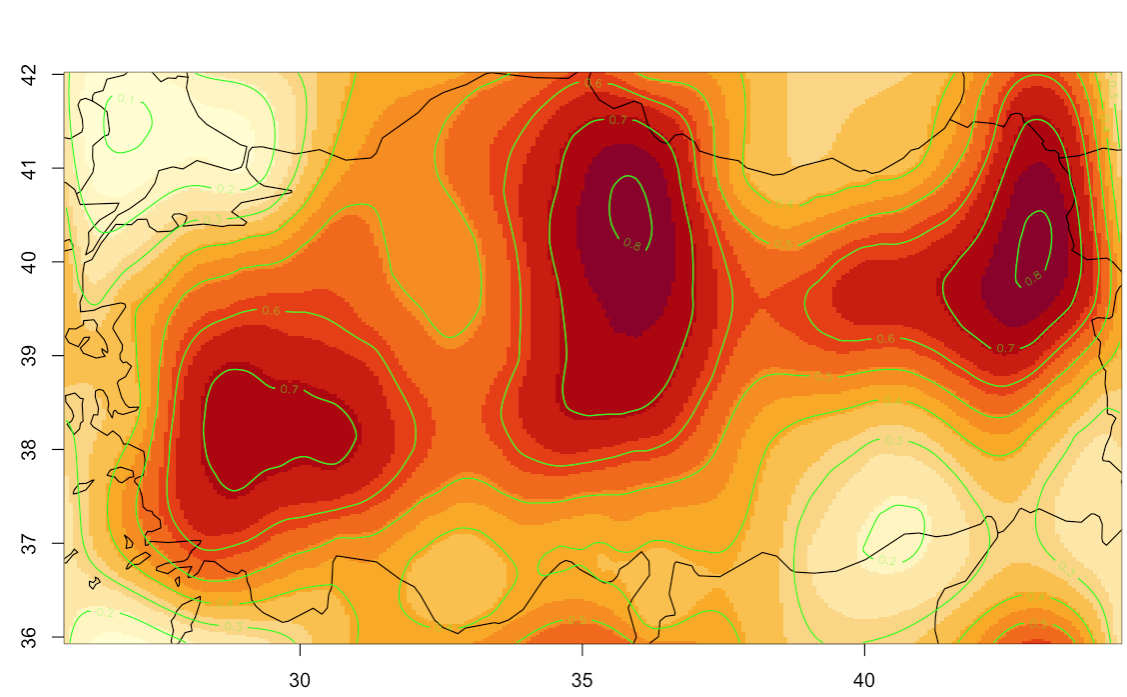}
            \caption[]{}%
        \end{subfigure}
        \hfill
        \begin{subfigure}[b]{0.475\textwidth}   
            \centering 
            \includegraphics[width=\textwidth]{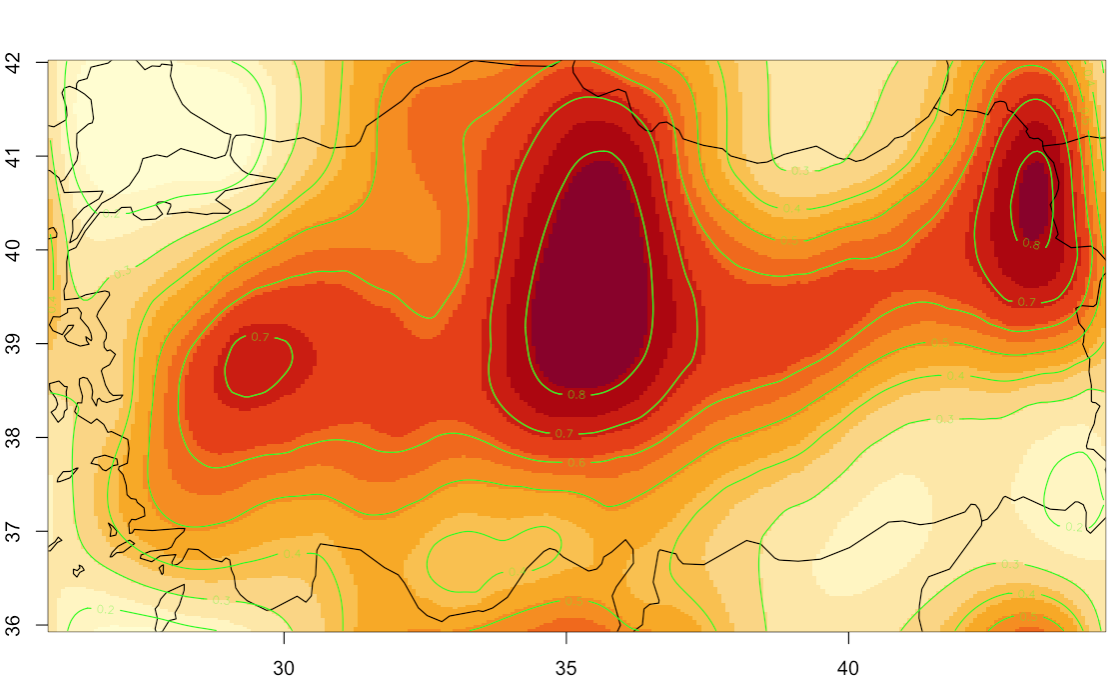}
            \caption[]{}%
        \end{subfigure}
        \caption{Disease map of FMD with serotype (a) A (b) O (c)  Asia-1 (d) all types}
        \label{fig:diseasemap}
    \end{figure*}

Due to low cross-immunity between FMD serotypes, we apply SDSM to type A, O and Asia-1 separately and generate unit-level susceptibility maps in Figure \ref{fig:diseasemap}. While the time series of outbreaks for each serotype follows different patterns (Figure \ref{fig:strain_ts}), we found relatively consistent maps of susceptibility across all three serotypes. These maps showed clusters of high susceptibility in the northeastern region, a region associated with the regional livestock industry  which has long distance connections to central Anatolia and Western Turkey \citep{yildiz2006livestock}. 


Thrace region remains a low risk region which has been  declared FMD free in 2010. In Central Anatolia region, a high risk cluster is also shared across different types of serotypes. The fact that the susceptibility  appear to be generally invariant to serotype is helpful for understanding the consistency in susceptibility across types, and may therefore be relevant for general  public health policy.

\subsection{England and Wales Measles Data}\label{england_wales}

For each week between 1960 and 1966, each municipality in England and Wales is either infected or susceptible based on whether any measles cases were reported in that municipality during that week.
Using the heuristic in Section \ref{dependence_test}, we find the independent model (ISM) should be sufficient to explain the susceptibility across cities in England and Wales. 

Our susceptibility estimates are highly correlated with population size at municipality level (we look at the relationship on log scale). Similar relationships between disease risk and population have been detected by prior analyses \citep{keeling1997disease, bjornstad2008hazards}. 

 \begin{figure*}[h!]
            \centering 
            \includegraphics[width=0.5\textwidth]{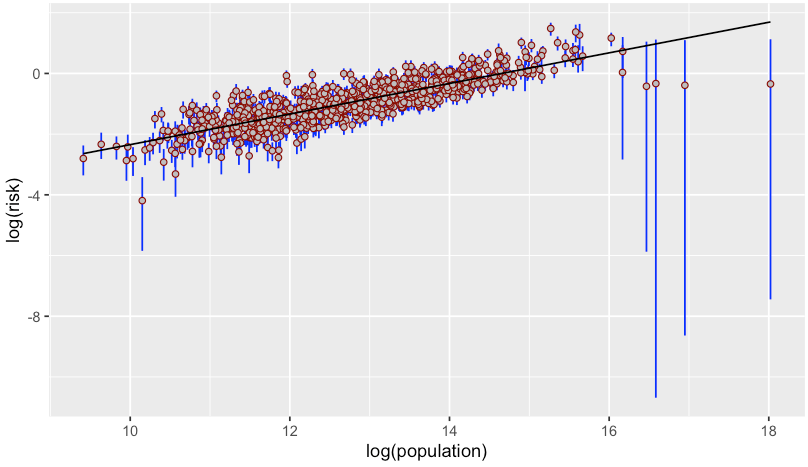}
        \caption{log of population size vs log risk of susceptibility with 95\% credible interval}
        \label{fig:diseasemap_time}
    \end{figure*}
    
  In Figure \ref{fig:diseasemap_time}, we observe strong positive correlation between log susceptibility risk and log population, with five outlying points with large population and wide credible interval for susceptibility risk. The five points correspond to London, Birmingham, Liverpool, Manchester and Sheffield which, due to their size,  rarely have weeks with zero measles cases. Our methods are most useful for describing the transmission behavior on units in susceptible status, and thus rely on the location being susceptible at many time points. As a result, for large cities that remain infected most of the weeks, we have limited information for fitting our model. Hence, the uncertainties for estimated susceptibility tends to be high for these cities. 

\begin{table}[hbt!]
\begin{tabular}{|l|l|l|l|l|l|}
\hline
 & Time steps & Level of dependence & ISM &  SDSM & SDSM PICAR\\ \hline
 exp 1& 100 & independent(0.10)& \textbf{0.703}& 0.461 & 0.133(0.13)  \\ \hline
 exp 2&  150& independent&  \textbf{0.808} & 0.678 & 0.135 \\ \hline
 exp 3&  200& independent & \textbf{0.868}  & 0.765 & 0.145 \\ \hline
 exp 4&  100& low & 0.625(0.09) & \textbf{0.772} & 0.735(0.016) \\ \hline
 exp 5&  150& low & 0.641 & \textbf{0.844} & 0.734 \\ \hline
 exp 6&  200& low & 0.696 & \textbf{0.861} & 0.773 \\ \hline
 exp 7& 100 &  medium & 0.528 & 0.735 & \textbf{0.836} \\ \hline
 exp 8& 150 & medium & 0.666 & 0.792 & \textbf{0.841} \\ \hline
 exp 9&  200& medium & 0.758 & 0.842  & \textbf{0.843} \\ \hline
 exp 10& 100 & high & 0.748 & \textbf{0.841} & 0.830 \\ \hline
exp 11 & 150 &  high & 0.805 & \textbf{0.903} & 0.871 \\ \hline
 exp 12& 200 & high  & 0.842 & \textbf{0.910} &  0.887\\ \hline
\end{tabular}
\end{table}

\section{Discussion}
\label{sec:conc}
In this paper, we proposed a general infectious disease model that accounts for both space-time transmission and the intrinsic susceptibility of that region to an outbreak. This susceptibility may arise from unknown or unmeasured factors such as population, health access, and contact pattern. We infer it only using the space-time records of outbreaks.   Both ISM and SDSM provide flexibility in unit-level susceptibility that varies spatially. While ISM has the advantage of providing a reliable risk map quickly, SDSM provides more accurate results when it is important to include spatial dependence in the susceptibilities. 
 Our simulation studies show that in cases with spatially independent susceptibility risk, ISM outperforms the others and in cases with spatially dependent susceptibility risk, SDSM outperforms ISM. At the same time, SDSM and SDSM-PICAR provide comparable performance in recovering the ground truth while utilizing PICAR improves computation speed and stability. When applying our method to the pre-vaccine era England and Wales measles surveillance data, we successfully detected the strong correlation between the inferred underlying disease risk with population. This result agrees with findings from prior study, without using information about population in the model. For Foot-and-mouth disease in Turkey, we detected three regions that have high risk of susceptibility. The three regions are in the northeast, center and western Turkey. These results are consistent across all three serotypes. 

 It is worth noting that our model relies on each location being in the susceptible state for at least one time unit. If the location almost never enters the susceptible state, it is hard to carry out reliable inference on the susceptibility parameter.  This requirement is fully met by our FMD data. For measles data, it is met for all but 5 cities (add the city names). This is why, as shown in  Section \ref{england_wales}, the estimates for  the four big cities  where the cities remain infected through most of the observation period, display large uncertainties. Hence, in addition to being aware of this potential issue, users need to choose the level of aggregating data wisely to ensure that there is enough information to infer susceptibilities. 

In the Turkey data set, where different serotypes have little cross-immunity, it suffices to subset data based on different serotypes and apply the model separately to each serotype. In future applications to disease with significant cross-immunity across serotypes, we would need to extend the current model to include these interactions and study the transmission behavior jointly.

\newpage
\section*{Acknowledgement}
The authors were partially supported by NSF-NIH-NIFA Ecology and Evolution of Infectious Disease award DEB 1911962 and NSF/BBSRC with ocde BB/T004312/1.

\bibliographystyle{style.bst}
\bibliography{bbb.bib}

\end{document}